\documentclass{article}
\usepackage{spconf,amsmath,graphicx,hyperref}
\usepackage{xcolor}
\usepackage{amssymb}
\usepackage{bbding}   
\usepackage{array}
\usepackage{booktabs}
\usepackage{float}
\usepackage{placeins}

\title{SSIMuse: Assessing Data Replication in Symbolic Music via an Adapted Structural Similarity Index Measure}
\name{Shulei Ji$^{1}$, Zihao Wang$^{2}$, Le Ma$^{1}$, Jiaxing Yu$^{1}$}
\address{$^{1}$Zhejiang University, $^{2}$Shandong University}
\begin{document}
	\ninept
	\maketitle
	\begin{abstract}
		AI-generated music may inadvertently replicate training samples and raise plagiarism concerns. Similarity measures can quantify such replication, thereby offering supervision and guidance for music generation models. Existing similarity measures for symbolic music mainly target monophonic melodies and provide limited support for complex musical textures, which restricts their ability to assess replication in polyphonic music. To address this limitation, we explore a simple cross-domain idea and propose \textbf{SSIMuse}, the first adaptation of the Structural Similarity Index Measure (SSIM) from images to symbolic music. SSIMuse is based on an image-like representation of symbolic music using binary piano rolls. It reinterprets the SSIM components as note density consistency and local note event overlap, and searches for replication under temporal shifts, pitch transpositions, and musically meaningful duration ratios. Controlled experiments on polyphonic and melodic datasets show that SSIMuse detects replication of at least one bar under exact and time-scaled conditions. Real-case melody retrieval further shows that it remains competitive with one symbolic plagiarism detector and two audio embedding models.
		The code is available at \url{https://github.com/Tayjsl97/SSIMuse}.
	\end{abstract}
	\section{Introduction}
	\label{sec:introduction}
	Despite impressive advances in music generation, relatively little effort has been devoted to evaluating originality or detecting plagiarism \cite{27,28}. This gap poses risks of copyright infringement, especially for musicians integrating such models into their workflows. Early studies show that Music Transformer \cite{25} tends to repeat input pieces due to overfitting in later training epochs \cite{1}, while diffusion models also pose high risks of data memorization \cite{12,26}. MusicLDM \cite{10} explicitly acknowledges these risks and proposes beat-synchronous mix-up strategies for data augmentation \cite{9}.
	Consequently, assessing replication and similarity in music has become an important research challenge, with applications extending beyond plagiarism detection to retrieval and recommendation. 
	
	Musical similarity can be assessed using symbolic or audio representations \cite{5,6,7,2,4}. This work focuses on detecting replication of note-level content
	in symbolic music, specifically the pitches and durations encoded in MIDI. Previous melody-oriented approaches typically require a monophonic note sequence, limiting their direct application to music containing simultaneous notes. We therefore investigate a similarity measure that directly compares note-level content without requiring melody extraction, supporting both melodic and accompaniment inputs.
	
	We propose \textbf{SSIMuse}, a training-free similarity measure
	for detecting replication in symbolic music, which transfers SSIM \cite{21} from image analysis to symbolic music. The image-like piano roll representation provides a natural bridge between these two domains. SSIMuse reinterprets its components in terms of note
	density consistency and local note overlap, and searches
	for replication under temporal shifts, pitch transposition,
	and uniform duration scaling. Experiments on controlled replication demonstrate that SSIMuse detects both exact and time-scaled copies, while retrieval experiments on real-world plagiarism cases show competitive performance against the previous
	music plagiarism detector and audio embedding models.
	\section{Related Work}
	\label{sec:related_work}
	Prior work on musical similarity measurement and plagiarism
	detection can be broadly divided into audio- and symbol-based methods.
	
	\textbf{Audio-based}. Audio fingerprints are widely applied for plagiarism detection, providing compact representations that identify audio fragments. Approaches include frequency-domain peak extraction with segment matching \cite{13} and fuzzy clustering for quantitative similarity assessment \cite{17}. Separately, neural network-based models learn music similarity from curated datasets, such as Siamese CNNs with triplet loss \cite{14} or MERT-based \cite{15} triplet networks for melodic similarity \cite{2}, though their out-of-domain generalization remains uncertain. Model-agnostic frameworks \cite{9} evaluate audio-based metrics like CLAP \cite{18} and FAD \cite{20} in exact data replication. Even measures originally designed for music retrieval \cite{16} can indicate potential plagiarism through high similarity scores.
	
	\textbf{Symbol-based}. Symbolic music can be represented as sequences or images for similarity computation. Sequence-based methods include BMM-Det \cite{7}, which uses bipartite graph matching, and an originality assessment framework based on geometric hashing \cite{1}. MPD-S \cite{4} converts MIDI to grayscale images and applies a Siamese CNN to melody similarity. Other methods adapt text similarity and clustering to discrete musical tokens \cite{5,6}. SSIMuse is based on the image-like piano roll representation and requires no model training.
	\vspace{-1.1em}
	\begin{figure*}[t]
		\centering
		\includegraphics[width=.92\linewidth]{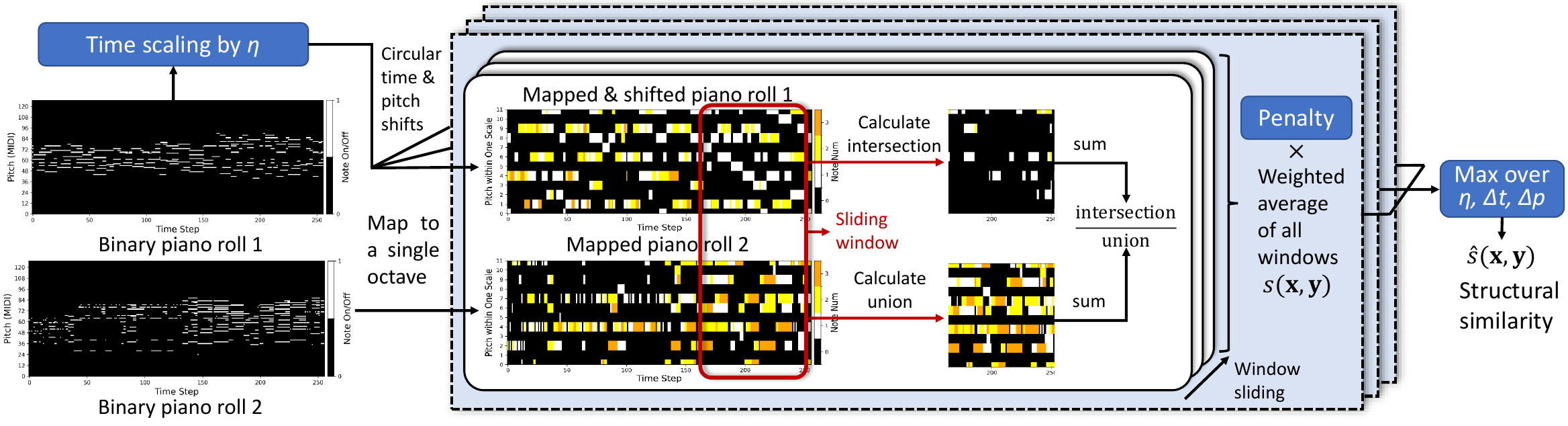}
		\vspace{-1em}
		\caption{The entire computation pipeline of $s_B(\mathbf{x},\mathbf{y})$, including duration scaling, circular shift alignment, and window aggregation.}
		\vspace{-0.5em}
		\label{fig:2}
	\end{figure*}
	\section{SSIMuse}
	\label{sec:ssimuse}
	\subsection{Preliminary}
	The structural similarity index measure (SSIM) \cite{21} quantifies the similarity between two images and is computed as follows:
	\begin{equation}
		\setlength{\abovedisplayskip}{5pt}
		\setlength{\belowdisplayskip}{5pt}
		\mathrm{SSIM}\left(\mathbf{x},\mathbf{y}\right)=\left[l\left(\mathbf{x},\mathbf{y}\right)\right]^{\alpha}\cdot\left[c\left(\mathbf{x},\mathbf{y}\right)\right]^{\beta}\cdot\left[s\left(\mathbf{x},\mathbf{y}\right)\right]^{\gamma}
	\end{equation}
	\begin{equation}
		\setlength{\abovedisplayskip}{5pt}
		\setlength{\belowdisplayskip}{5pt}
		l(\mathbf{x},\mathbf{y})=\frac{2\mu_{x}\mu_{y}+C_{1}}{\mu_{x}^{2}+\mu_{y}^{2}+C_{1}}
	\end{equation}
	\begin{equation}
		c(\mathbf{x},\mathbf{y})=\frac{2\sigma_{x}\sigma_{y}+C_{2}}{\sigma_{x}^{2}+\sigma_{y}^{2}+C_{2}}
	\end{equation}
	\begin{equation}
		s(\mathbf{x},\mathbf{y})={\frac{\sigma_{x y}+C_{3}}{\sigma_{x}\sigma_{y}+C_{3}}} \label{eq:str}
	\end{equation}
	\noindent where $\mu_{k},\sigma_{k}\ (k\in\{\mathbf{x},\mathbf{y}\})$ are the mean and standard deviation of image $k$. $l(\cdot)$, $c(\cdot)$, and $s(\cdot)$ are the luminance, contrast, and structure comparison functions, measuring mean consistency, standard deviation consistency, and structural similarity via covariance, respectively. $\alpha$, $\beta$, and $\gamma$ are parameters adjusting the relative importance of these components. $C_{1},C_{2},C_{3}$ are constant parameters. Following original SSIM, $\alpha=\beta=\gamma=1$ and $C_{3}=C_{2}/2$.
	
	SSIM evaluates the similarity within local windows and mean SSIM (MSSIM) evaluates the whole image, as follows:
	\begin{equation}
		\setlength{\abovedisplayskip}{3pt}
		\setlength{\belowdisplayskip}{3pt}
		\mathrm{MSSIM}\left(\mathbf{X},\mathbf{Y}\right)=\frac{1}{M}\sum_{j=1}^M{SSIM\left(x_j,y_j\right)} \label{eq:mssim}
	\end{equation}
	where $x_j$ and $y_j$ are the contents in the $j$th window of the two images, respectively, and $M$ is the number of local windows. In this paper, MSSIM is used to compute the similarity of two piano roll representations corresponding to two musical clips.

	\subsection{SSIMuse}
	\label{sec:B}
	Symbolic music is first represented as a binary piano roll $G \in \{0,1\}^{T \times P}$, where $T$ is the number of time steps and $P=128$ is the number of MIDI pitches. We quantize each quarter note into four time steps, yielding 16 steps per bar in 4/4 meter. Each element $G_{ij}$ equals 1 if pitch $j$ is active at time step $i$, and 0 otherwise.
	We then adapt SSIM to binary piano rolls by reinterpreting its three comparison functions as follows:
	
	1) $l_B(\mathbf{x},\mathbf{y})$: measures the note density consistency between musical excerpts as the mean of a binary PR reflects note density. 
	
	2) $c_B(\mathbf{x},\mathbf{y})$: becomes invalid for the binary PRs. Two binary PRs with the same mean necessarily have the same standard deviation, which cannot reflect fluctuations. Moreover, when the sum of two means equals one (e.g., in special cases where the positions of 1s and 0s are inverted), the standard deviations remain identical, leading to ambiguity. Therefore, this function is omitted when computing similarity between binary PRs.  
	
	3) $s_B(\mathbf{x},\mathbf{y})$: According to Equation~(\ref{eq:str}), with the standard deviation invalid, only the covariance in the numerator of the structure comparison function is retained, computed as:
	\begin{equation}
		\setlength{\abovedisplayskip}{3pt}
		\setlength{\belowdisplayskip}{3pt}
		s_B(\mathbf{x},\mathbf{y})=\sigma_{x y} = \frac{1}{N-1} \sum_{i=1}^{N} (x_i - \mu_x)(y_i - \mu_y)
	\end{equation}
	Because binary PRs contain many zeros for silences, the covariance tends to be inflated, making it difficult to distinguish between PRs. To address this, $s_B(\mathbf{x},\mathbf{y})$ is computed only at positions where at least one note occurs in either PR. Moreover, unlike continuous signals that require mean subtraction to remove global offsets, binary PRs have no such concept. Subtracting the mean not only dilutes the co-occurrence of note-on events but also introduces density-related bias. Thus, for binary PRs, we discard mean subtraction and focus on the co-occurrence of note-on events across positions to capture note-event synchrony, simplifying $s_B(\mathbf{x},\mathbf{y})$ into:
	\begin{equation}
		\setlength{\abovedisplayskip}{3pt}
		\setlength{\belowdisplayskip}{3pt}
		s_B(\mathbf{x},\mathbf{y})=\frac{1}{N^{\prime}}\sum_{i=1}^{N^{\prime}}x_{i}\cdot y_{i}\equiv\text{Jaccard}(\mathbf{x},\mathbf{y})=\frac{|\mathbf{x}\cap\mathbf{y}|}{|\mathbf{x}\cup\mathbf{y}|}
	\end{equation}
	where $N^{\prime}$ is the number of positions where at least one note occurs. At these positions, $s_B(\mathbf{x},\mathbf{y})$ computes the average of element-wise products, equivalent to the intersection over the union of note-on events, i.e., the Jaccard index. As a classic similarity metric for binary data, the Jaccard index provides an intuitive and precise measure of note-event overlap for binary PRs.
	
	 Based on the Jaccard index, we introduce three refinements for more accurate similarity computation, as shown in Fig.~\ref{fig:2}.

	\noindent\textbf{i) Octave folding.} To match copied note patterns across octave changes, we map the PR into 12 pitch classes:
	\begin{equation}
		\setlength{\abovedisplayskip}{3pt}
		\setlength{\belowdisplayskip}{3pt}
		\hat{G}(i,k) = \sum_{j \bmod 12 = k} G(i,j), \quad k \in [0,11], \; j \in [0,127]
	\end{equation}
	The summation retains the number of active pitches in each pitch class while discarding their octave positions. After octave folding, the Jaccard score becomes the sum of element-wise minima divided by the sum of element-wise maxima.

	\noindent\textbf{ii) Joint scale and shift alignment.} Copied excerpts may occur at different temporal positions, be transposed in pitch, or be uniformly scaled in duration. To account for these transformations, we first scale the entire candidate $\mathbf{y}$ by $\eta$ to obtain $\mathbf{y}_{\eta}=\operatorname{Resize}(\mathbf{y},\eta)$, and then circularly shift it by $\Delta t$ time steps and $\Delta p$ pitch classes. The integer search ranges are $|\Delta t|\leq\lfloor T_{\eta}/2\rfloor$ and $-5\leq\Delta p\leq6$, where $T_{\eta}=\max(T_x,\operatorname{round}(\eta T_y))$. The scale factor is selected from $\eta\in\mathcal{H}=\{1/16,1/8,1/4,1/3,3/8,1/2,2/3,3/4,1,4/3,3/2,2,\\8/3,3,4,8,16\}$, which contains powers of two and triplet-related duration ratios. Every ratio is paired with its reciprocal so that compression and expansion are searched symmetrically. Setting $\mathcal{H}=\{1\}$ reduces the method to shift-only matching. Each scaled candidate is computed once and reused for all shifts. 

	To penalize larger deviations from the original temporal
	alignment and duration scale, we introduce temporal and
	scale penalties into the similarity score. The resulting
	structural score is
	\begin{equation}
		\setlength{\abovedisplayskip}{3pt}
		\setlength{\belowdisplayskip}{3pt}
		\hat{s}_B=
		\max_{(\Delta t,\Delta p,\eta)\in\mathcal{T}\times\mathcal{P}\times\mathcal{H}}
		{s}_{B,(\eta,\Delta t,\Delta p)}\cdot\pi_t(\Delta t)\cdot\pi_{\mathrm{sc}}(\eta)
		\label{eq:msshift}
	\end{equation}
	where the temporal and scale penalties are
	\begin{equation}
		\setlength{\abovedisplayskip}{3pt}
		\setlength{\belowdisplayskip}{3pt}
		\begin{aligned}
		\pi_t(\Delta t)&=\max\!\left(0,1-\lambda |\Delta t|/T_{\eta}\right),\\
		\pi_{\mathrm{sc}}(\eta)&=\max\!\left(0,1-\lambda'|\log_2\eta|\right).
		\end{aligned}
		\label{penality}
	\end{equation}
	Here, $\lambda$ and $\lambda'$ control the temporal and scale penalties, respectively.
	
	\noindent\textbf{iii) Window aggregation.} A copied excerpt may occupy only a small portion of the music, so averaging similarity over the entire sequence can dilute local replication evidence. To emphasize local matches, we divide the aligned PRs into $M$ temporal windows for each $(\eta,\Delta t,\Delta p)$ and aggregate their window-level Jaccard scores using similarity-adaptive weights in the controlled-replication experiments:
	\begin{equation}
		\setlength{\abovedisplayskip}{3pt}
		\setlength{\belowdisplayskip}{3pt}
		{s}_{B,(\eta,\Delta t,\Delta p)}
		=\frac{\sum_{i=1}^{M}w^i{s}^i_{B,(\eta,\Delta t,\Delta p)}}
		{\sum_{i=1}^{M}w^i}
		\label{eq:window_aggregate}
	\end{equation}
	where $s_B^i$ is the structural similarity in the $i$-th window. Each window spans all pitch classes and slides only along time, weighted by a window weighting function $w(s)$ to amplify highly overlapping segments and suppress dissimilar ones. Setting $w(s)\equiv1$ reduces Equation~(\ref{eq:window_aggregate}) to the classical MSSIM average in Equation~(\ref{eq:mssim}).

	The exhaustive shift search above is accelerated using a two-dimensional fast Fourier transform (FFT). Let $A$ denote the octave-folded PR of $\mathbf{x}$ and $B_{\eta}$ the octave-folded PR of the scaled candidate $\mathbf{y}_{\eta}$. Because these PRs contain nonnegative integer counts, we decompose them into threshold layers: $A_i^{(k)}$ is $\mathbb{I}[A\geq k]$ inside the $i$-th temporal window and zero elsewhere, while $B_{\eta}^{(k)}=\mathbb{I}[B_{\eta}\geq k]$ retains the thresholded values over the entire PR. The intersection counts for all shifts are computed simultaneously as
	\begin{equation}
		I_{i,\eta}(\Delta t,\Delta p)=\sum_{k=1}^{K}\mathcal{F}^{-1}\!\left(\overline{\mathcal{F}(A_i^{(k)})}\odot\mathcal{F}(B_{\eta}^{(k)})\right)[\Delta t,\Delta p]
		\label{eq:fft}
	\end{equation}
	where $K$ is the maximum folded count, $\mathcal{F}$ is the 2-D FFT, $\overline{\mathcal{F}}$ denotes complex conjugation, and $\odot$ is element-wise multiplication. Let $B_{\eta,i}^{(\Delta t,\Delta p)}$ denote the shifted candidate restricted to this window. Now, the corresponding union and window-level Jaccard score are
	\begin{equation}
		U_{i,\eta}(\Delta t,\Delta p)
		=\sum A_i+\sum B_{\eta,i}^{(\Delta t,\Delta p)}-I_{i,\eta}(\Delta t,\Delta p)
	\end{equation}
	\begin{equation}
		s^i_{B,(\eta,\Delta t,\Delta p)}
		=\frac{I_{i,\eta}(\Delta t,\Delta p)}{U_{i,\eta}(\Delta t,\Delta p)}
		\label{eq:fft_jaccard}
	\end{equation}
	The score is set to zero when the union is zero. 
	\section{Experiments}
	\label{sec:experiments}
	\subsection{Datasets}
	\label{sec:datasets}
	For controlled-replication experiments, we use two source datasets. The first is the \textbf{Ailabs} \cite{23}, which contains 1,747 piano cover performances of pop songs in 4/4 meter. The second is the \textbf{POP909} \cite{24}, comprising 909 professional piano arrangements of pop songs. We retain only samples with quarter-note beats and convert them into 4/4 meter, yielding 899 samples. In addition, we extract melody lines from POP909 to construct the \textbf{POP909-melody} dataset. For real-case source retrieval, we use two additional datasets. The \textbf{Real-life Dataset} from~\cite{7} contains 29 original--plagiarized pairs. The \textbf{music-copyright dataset}~\cite{5} contains 18 adjudicated infringement pairs and 22 non-infringement pairs. One case contains no pitched notes and is excluded from the candidate set, leaving 39 candidates. 
	\subsection{Controlled-Replication Data Construction}
	Following \cite{9}, we construct controlled replication data by placing either unchanged or time-scaled reference excerpts into unrelated mixture samples. We refer to $\eta_{\mathrm{gt}}=1$ as \textbf{unscaled replication} and $\eta_{\mathrm{gt}}\neq1$ as \textbf{time-scaled replication}. Each dataset is segmented into 256-time-step (16-bar) clips and divided into disjoint \textbf{reference} and \textbf{mixture} sets of 400 clips each. We use 4,000 pairs sampled without replacement from the $400\times400$ reference--mixture combinations as the no-replication baseline. We consider four levels of replication: 1 bar (6.25\%), 2 bars (12.5\%), 4 bars (25\%), and 8 bars (50\%) before scaling. For every reference and every valid scale $\eta_{\mathrm{gt}}$, one randomly selected $b$-bar ($b\in\{1,2,4,8\}$) excerpt is scaled, a mixture clip is randomly selected, and the scaled excerpt replaces the mixture content at a randomly selected position to form the \textbf{target} clip. Note that if the scaled excerpt ends with a fractional time step, we
	allocate one additional time step and set its value to the
	remaining fraction. The scaled excerpt length is restricted to between 4 and 256 time steps. There are 15 valid scales for $b=1,2,4$ and 12 for $b=8$. Thus, the four replication levels contain 6,000, 6,000, 6,000, and 4,800 target--reference pairs, respectively.

	\subsection{Compared Methods and Evaluation Settings}
	We compare SSIMuse with one symbolic and two audio-based methods. BMM-Det \cite{7} is a bipartite matching detector designed for monophonic melodies. For a polyphonic piano roll without a specified melodic line, we retain the highest active pitch at each non-silent time step and merge consecutive frames with the same pitch. CLAP \cite{18} and MERT \cite{15} operate on audio. We render every MIDI pair with the same piano soundfont and then extract the audio embeddings to compute cosine similarity.
	
	We set $\lambda=0.5$ and $\lambda'=0.125$ in Equation~(\ref{penality}). For experiments on the controlled-replication data, all methods are evaluated on identical target--reference pairs and baseline pairs. We use the linear weighting function $w(s)=s$ in Equation~(\ref{eq:window_aggregate}), with both the window size and hop size set to 16 time steps (one bar). For real-case retrieval, SSIMuse uses a two-level comparison scheme. Each melody is divided into 8-bar retrieval segments with a 6-bar hop for song-level matching. Within each segment pair, structural similarity is computed using 1-bar local windows with a 1-bar hop and uniform weighting, i.e., $w(s)\equiv1$.
	\begin{figure*}[t]
		\centering
		\includegraphics[width=\textwidth]{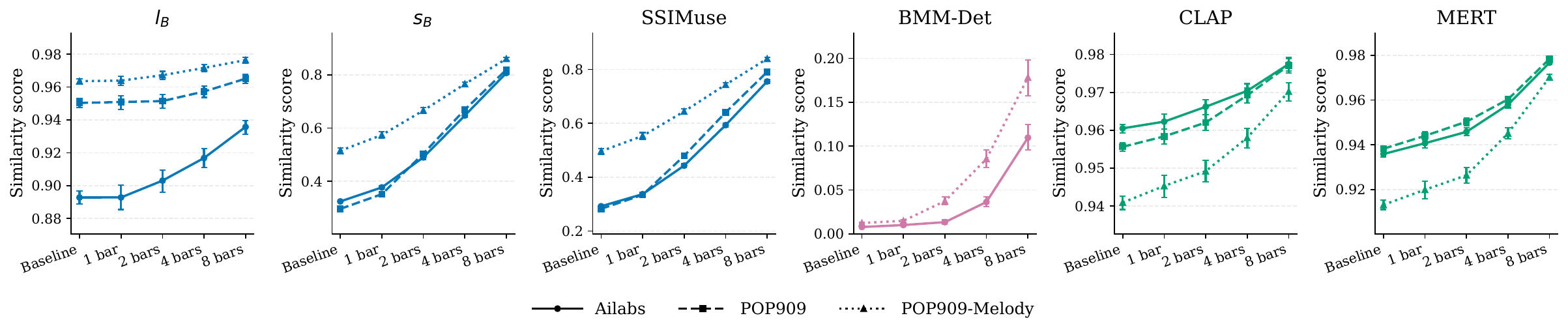}
		\vspace{-2.2em}
		\caption{Results for unscaled replication ($\eta_{\mathrm{gt}}=1$) on Ailabs, POP909, and POP909-melody. Points show means and error bars show 95\% confidence intervals. BMM-Det's POP909 result is omitted because POP909-melody provides the corresponding monophonic setting.}
		\label{fig:unscaled_replication}
		\vspace{-1em}
	\end{figure*}
	\begin{figure}[t]
		\centering
		\includegraphics[width=\columnwidth]{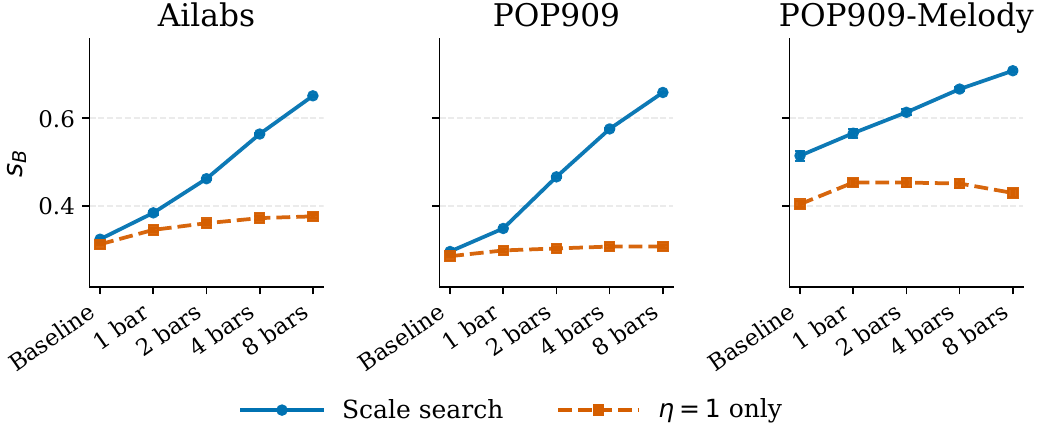}
		\vspace{-2.2em}
		\caption{$s_B$ on time-scaled replication ($\eta_{\mathrm{gt}}\neq1$), with and without scale search.}
		\label{fig:scale_search_ablation}
		\vspace{-1em}
	\end{figure}
	
	\begin{figure}[t]
		\centering
		\includegraphics[width=\columnwidth]{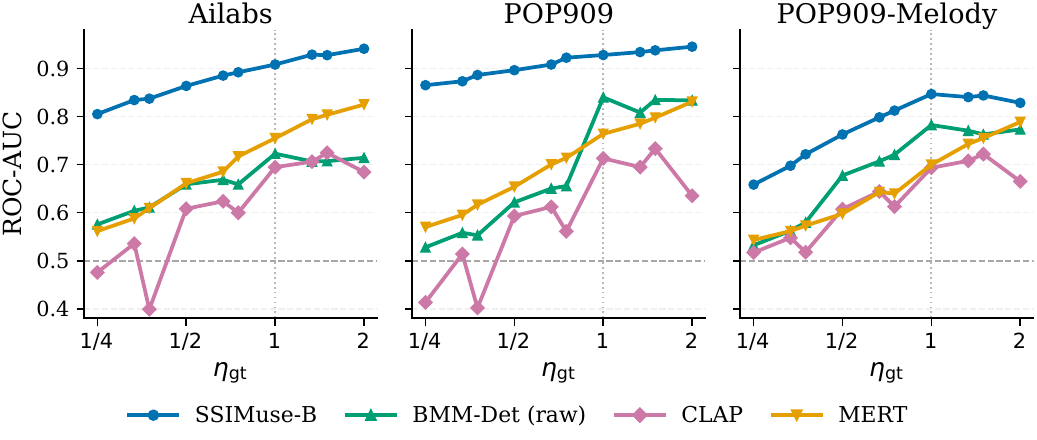}
		\vspace{-2.2em}
		\caption{ROC-AUC across injected time-scale ratios. Each point uses the same 4,000 baseline pairs as negatives and 1,600 controlled copies as positives, evenly distributed across 1, 2, 4, and 8 bars. }
		\label{fig:scale_robustness}
		\vspace{-1em}
	\end{figure}
	\subsection{Results on Unscaled Replication}
	Figure~\ref{fig:unscaled_replication} compares all metrics on unscaled replicated excerpts. For SSIMuse, the density term $l_B$ increases only moderately because inserting a local excerpt changes global note density relatively little. In contrast, $s_B$ and SSIMuse rise consistently from the no-replication baseline to 8-bar replication on all three datasets. POP909-melody generally yields higher scores because its sparse texture produces less competing note content after alignment. BMM-Det exhibits a clear response for longer copied excerpts, while its separation between the baseline and 1-bar replication is limited. The scores of the audio-based methods CLAP and MERT also increase with replication duration, but their scores occupy a narrow, high-similarity range, especially from the baseline to short replications. Overall, all methods distinguish between different numbers of replicated bars.
	\subsection{Results on Time-Scaled Replication}
	Figure~\ref{fig:scale_search_ablation} compares $s_B$ with scale search ($\eta\in\mathcal{H}$) and without it ($\eta=1$) on genuinely time-scaled samples. Scale search produces a clear increase with replication duration on all three datasets, whereas fixing $\eta=1$ yields substantially weaker or non-monotonic responses, especially for longer excerpts. This confirms that temporal-shift alignment alone cannot
	reliably recover overlap after duration scaling, highlighting
	the need for explicit scale search.

	\subsection{Robustness across Time-Scale Ratios}
	To assess replication detection across scaling ratios,
	Figure~\ref{fig:scale_robustness} compares four methods
	at ten ratios using ROC-AUC, which measures their ability
	to rank controlled-copy pairs above no-replication pairs.
	SSIMuse performs best at all ratios on three datasets.
	BMM-Det generally improves with increasing ratios but
	shows fluctuations.
	MERT is the strongest audio baseline and improves
	consistently with increasing ratios. CLAP is sensitive
	to the ratio, with local drops at
	$\eta_{\mathrm{gt}}=3/8$ and $3/4$ and a decline
	at $2$ across all three datasets.
	
	\subsection{Computational Efficiency}

	\begin{figure}[t]
		\centering
		\includegraphics[width=\columnwidth]{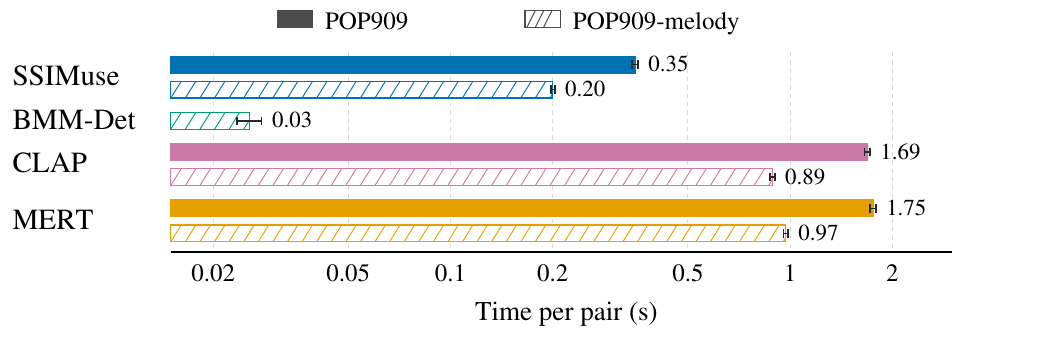}
		\vspace{-2em}
		\caption{Per-pair runtime averaged over 100 16-bar baseline pairs on one NVIDIA A800 GPU, with the horizontal axis on a logarithmic scale. For audio-based methods, the time includes MIDI rendering and embedding extraction but excludes model initialization.}
		\label{fig:runtime}
		\vspace{-1em}
	\end{figure}
	\begin{table}[t]
		\centering
		\caption{Top-1 accuracy of melody-based source retrieval on two
			real-case datasets.}
		\label{tab:real_plagiarism}
		{\normalsize
			\setlength{\tabcolsep}{5.0pt}
			\begin{tabular*}{\columnwidth}{@{\extracolsep{\fill}}lcc@{}}
				\toprule
				Method & \shortstack{29 queries / 29 candidates} & \shortstack{18 queries / 39 candidates} \\
				\midrule
				SSIMuse & 79.3\% (23/29) & \textbf{50.0\% (9/18)} \\
				BMM-Det & \textbf{82.8\% (24/29)} & 44.4\% (8/18) \\
				CLAP & 41.4\% (12/29) & 27.8\% (5/18) \\
				MERT & 55.2\% (16/29) & \textbf{50.0\% (9/18)} \\
				\bottomrule
		\end{tabular*}}
	\end{table}
	
	Figure~\ref{fig:runtime} compares per-pair latency under the same fixed-clip protocol, with all methods timed sequentially in one job on the same GPU. BMM-Det is the fastest, followed by SSIMuse, CLAP, and MERT. SSIMuse processes each pair in under half a second on average and is 4.4--5.0 times faster than the audio-based methods.
	\subsection{Melody Source Retrieval on Real Cases}
	Following the ranking-based evaluation in \cite{7}, we evaluate melody-based source retrieval on the two real-case datasets introduced in Section~\ref{sec:datasets}. Specifically, 
	SSIMuse uses the Hungarian algorithm to obtain a maximum-weight one-to-one matching between query and candidate segments, then retains the top three matched pairs. Their similarities are weighted by non-duplicated query coverage and normalized by the full query duration. Before ranking, each score is divided by the candidate original’s mean score across all queries.
	The Top-1 accuracy is reported in Table~\ref{tab:real_plagiarism}. On the 29-case dataset, BMM-Det obtains the highest Top-1 accuracy at 82.8\% (24/29), while SSIMuse follows at 79.3\% (23/29) and exceeds MERT and CLAP. On the more difficult 18-query/39-candidate setting, SSIMuse and MERT both reach 50.0\%, compared with 44.4\% for BMM-Det and 27.8\% for CLAP. These results show that SSIMuse is competitive with the strongest baselines.
	
	\section{Conclusion}
	\label{sec:conclusion}
	This paper adapts the Structural Similarity Index Measure from image analysis to symbolic music. SSIMuse represents music as an image-like binary piano roll and compares its note density and local note-event structure. It searches over temporal translation, pitch transposition, and musically meaningful duration ratios. Controlled experiments show that SSIMuse detects replication of at least one bar under exact and time-scaled conditions. In real-case melody retrieval, it identifies 23 of 29 sources in the Real-life Dataset and 9 of 18 copyright-case sources from 39 candidates.


	
	
	
	
	\small
	\bibliographystyle{IEEEbib}
	\bibliography{strings,refs}
	
\end{document}